\documentclass{ecai} 

\usepackage{latexsym}
\usepackage{amssymb}
\usepackage{amsmath}
\usepackage{amsthm}
\usepackage{booktabs}
\usepackage{enumitem}
\usepackage{graphicx}
\usepackage{color}
\usepackage[ruled,linesnumbered]{algorithm2e}
\usepackage{dsfont}
\usepackage{amsmath,amssymb,amsfonts}
\usepackage{algorithmic}
\usepackage{tabularx}
\usepackage{graphicx}
\usepackage{textcomp}
\usepackage{multirow,multicol}
\usepackage{booktabs}
\usepackage{nameref}
\usepackage{hyperref}

\newcommand{\BibTeX}{B\kern-.05em{\sc i\kern-.025em b}\kern-.08em\TeX}

\begin{document}


\begin{frontmatter}


\paperid{1056} 


\title{Deep Self-knowledge Distillation: A hierarchical supervised learning for coronary artery segmentation}


\author[A,B]{\fnms{Mingfeng}~\snm{Lin}\thanks{Corresponding Author. Email: 22920212204153@stu.xmu.edu.cn}}

\address[A]{Harbin Institute of Technology, Shenzhen, China}
\address[B]{Xiamen University, Xiamen, China}


\begin{abstract}
Coronary artery disease is a leading cause of mortality, underscoring the critical importance of precise diagnosis through X-ray angiography. Manual coronary artery segmentation from these images is time-consuming and inefficient, prompting the development of automated models. However, existing methods, whether rule-based or deep learning models, struggle with issues like poor performance and limited generalizability. Moreover, current knowledge distillation methods applied in this field have not fully exploited the hierarchical knowledge of the model, leading to certain information waste and insufficient enhancement of the model's performance capabilities for segmentation tasks. To address these issues, this paper introduces Deep Self-knowledge Distillation, a novel approach for coronary artery segmentation that leverages hierarchical outputs for supervision. By combining Deep Distribution Loss and Pixel-wise Self-knowledge Distillation Loss, our method enhances the student model's segmentation performance through a hierarchical learning strategy, effectively transferring knowledge from the teacher model. Our method combines a loosely constrained probabilistic distribution vector with tightly constrained pixel-wise supervision, providing dual regularization for the segmentation model while also enhancing its generalization and robustness. Extensive experiments on XCAD and DCA1 datasets demonstrate that our approach outperforms the dice coefficient, accuracy, sensitivity and IoU compared to other models in comparative evaluations.
\end{abstract}

\end{frontmatter}


\section{Introduction}\label{sec1}
Coronary artery disease stands as one of the leading causes of mortality globally, resulting in a significant loss of life \citep{golub2023major}. X-ray angiography, recognized as the gold standard for diagnosing coronary artery disease, is an indispensable tool in clinical diagnostics \citep{han2023coronary}. Consequently, the extraction of the coronary artery from X-ray angiography is essential for disease diagnosis. However, manual segmentation of the coronary arteries from X-ray angiography is not only time-consuming, but also represents a waste of limited medical resources. Consequently, numerous studies are currently focused on developing automated models for the segmentation of coronary arteries. Yet, these models are often hindered by limitations such as low contrast due to vessel stenosis and low spatial resolution, which impair their segmentation performance and prevent them from effectively aiding clinical physicians in diagnosis \citep{wang2023application}.

Traditional methods for coronary artery segmentation primarily rely on rule-based approaches, such as region growing methods, thresholding methods, and filtering methods. These techniques necessitate a series of predefined rules that are contingent upon prior expert knowledge and complex preprocessing procedure, thereby leading to models that lack generalizability and representational learning capabilities \citep{ma2021self}. Moreover, these rule-based methods have distinct shortcomings. For instance, region growing methods do not specifically account for the branching nature of coronary arteries, resulting in missed branches and disconnected vessels \citep{gharleghi2022towards}. Both thresholding and filtering methods are susceptible to missing branches due to the low contrast of coronary arteries in the original X-ray angiography images. In summary, conventional coronary artery segmentation methods are plagued with issues of suboptimal model performance and limited generalization capabilities \citep{wang2023application}.

With the advancement of deep learning, numerous deep learning-based methods have emerged in recent years. U-Net \citep{ronneberger2015u}, a milestone in the field of medical image segmentation, has achieved remarkable progress since its introduction in 2015. U-Net, which is an encoder-decoder architecture, effectively captures the contextual information of images while maintaining sensitivity to details through its symmetrical design of downsampling and upsampling. Currently, many literature has been dedicated to the development of U-Net's variants, such as Attention U-Net \citep{oktay2018attention}, ResUnet \citep{zhang2018road}, U-Net++ \citep{zhou2018unet++}, ResUnet++ \citep{jha2019resunet++}, and U-Net3+ \citep{huang2020unet}.

Knowledge distillation, a technique for extracting valuable information from a teacher model to guide the learning of a student model, has been widely applied across various deep learning technologies, including object classification, object detection, and machine translation \citep{hinton2015distilling}. Self-knowledge distillation, which involves extracting knowledge from the model itself to guide the student model's learning, effectively addresses the inconvenience of transferring knowledge between teachers and students from different sources \citep{gou2021knowledge}.

In the domain of image segmentation, the existing literature on self-knowledge distillation predominantly employs it as a tool during the training process to provide pixel-wise supervision for the final output \citep{ji2021refine}. However, in models with an encoder-decoder architecture, such as U-Net, each layer of the decoder contains feature maps with different levels of feature representation, which encapsulate the knowledge that the model has learned for a specific task. Focusing solely on pixel-wise supervision of the final output neglects this valuable information. Moreover, extracting additional knowledge from side outputs and utilizing it during the inference process incurs no additional computational cost.

Building on the aforementioned information, this paper introduces a deep self-knowledge distillation approach for coronary artery segmentation. This hierarchical learning strategy enables the teacher model to deeply convey useful information to the student model, thereby enhancing the segmentation performance of the student model. Deep self-knowledge distillation comprises two components: Deep distribution supervision and Pixel-wise supervision. Deep distribution supervision takes into account that the side feature maps have learned features at various granularities, thus employing a loosely constraint probabilistic distribution vector in contrast to the tight constraint of pixel-wise supervision to supervise both the teacher and student models. Pixel-wise supervision is applied to the final output, where the teacher model's predictions are linearly combined with the ground truth to form a soft target, supervising the student model's predictions and leading to improved generalization and robustness.

In summary, the main contributions of this paper can be encapsulated as follows:

\begin{itemize}
\item We propose a Deep Self-knowledge Distillation method that fully leverages the hierarchical features within the segmented models of the encoder-decoder architecture. By combining both loosely and tight constraints to optimize the model, we have enhanced the performance of coronary artery segmentation without incurring additional computational resources during the inference phase.
\item We introduce a more loosely constraint by transforming side outputs of different granularities as probabilistic distribution vectors, which introduces fewer inductive biases into the model optimization process, thereby making the model more robust.
\item Through comparative experiments, we have found that our proposed method achieves state-of-the-art results in terms of dice coefficient, accuracy, sensitivity, and IoU when compared to other models. Furthermore, additional ablation studies indicate that both components of Deep Self-knowledge Distillation individually contribute to the model's performance, with their combination yielding the greatest improvement.
\end{itemize}

The subsequent sections of this paper are organized as follows: Section "Related work" provides an overview of the current research on coronary artery segmentation and the application of self-knowledge distillation in segmentation tasks. Section "Methodology" begins with an introduction to some preliminary knowledge, followed by a detailed presentation of our proposed Deep Self-knowledge Distillation approach. Section "Experiment" first describes the datasets involved in this study and the experimental parameter settings, then proceeds to introduce comparative experiments and ablation studies, along with a qualitative and quantitative analysis of the comparative experiments. Finally, Section "Conclusion" summarizes the entire paper.

\section{Related Work}\label{sec2}
\subsection{Coronary Artery Segmentation}
Traditional rule-based methods encompass techniques such as region growing methods, thresholding methods, and filtering methods. \cite{tang2013segmentation} initially utilized multi-scale filtering to identify vessel and boundary regions, followed by region growing to detect the aorta and coronary arteries. \cite{hung2013automated} employed region growing based on vesselness and intensity to generate segmentation and applied discrete wavelet transform for post-processing to enhance the segmentation results. 
\cite{wang2010automated} applied a vesselness filter for image pre-processing, then used a mesh contraction algorithm to obtain the centerlines of the segmented coronary arteries, and finally employed an adaptive threshold to achieve coronary artery segmentation. \cite{khan2020hybrid} designed efficient denoising filters for image pre-processing to enhance segmentation effects, followed by a threshold-based method with a vessel location map to obtain the segmentation mask, and concluded with logical operations for post-processing to reinforce the vessel structure.

Recently, deep learning methods based on convolutional neural networks (CNNs) have made significant advancements in the field of coronary artery segmentation. \cite{fan2019accurate} proposed the Octave UNet, an encoder-decoder architecture that learns hierarchical multi-frequency features through octave convolution to better represent the representations of narrow vessels. \cite{li2022automatic} first extracted the region of interest (ROI) using U-Net from the original images, then input the ROI and clinical parameters into a 3DNet for the diagnosis of coronary artery disease. \cite{song2022automatic} integrated dense blocks into 3D-UNet to learn rich and representative features from coronary artery images and introduced a Gaussian weighting method for post-processing to enhance the model's segmentation performance. \cite{jiang2024ori} introduced a U-Net-like network named Ori-Net, which incorporates the topological prior of the coronary artery into the segmentation model and employs multi-task learning to learn segmentation, radius, and orientation, thereby improving the connectivity of segmented vessels.

\subsection{Self-Knowledge Distillation in Segmentation Task}
Most existing image segmentation works with self-knowledge distillation do not tailor the distillation process to specific tasks, but rather use it as a tool during training, with the objective of extracting effective knowledge from the teacher model to guide the optimization process of the student model for better segmentation results \citep{shen2023expert,park2022semantic,zhang2024self,zheng2023self}. Furthermore, many existing methods do not consider the varying granularity of features learned at different levels of feature extraction. Moreover, direct tight constraints on pixel-wise supervision can cause the network to be biased in the early stages of self-knowledge distillation, leading to poor segmentation outcomes.

\cite{shen2023expert} proposed a semi-supervised framework called EXP-Net, which enhances knowledge distillation by introducing an expert network distinct from student and teacher networks. \cite{park2022semantic} proposed a pixel-wise adaptive label smoothing to strengthen the generalization capability during the self-knowledge distillation process, supervising the softened label and the student model's output at the final output position of the model. \cite{zhang2024self} introduced self-knowledge distillation in the RGB-D Mirror Segmentation task, optimizing the pixel-wise supervision process with a combination of Kullback–Leibler divergence and dice loss. \cite{zheng2023self} proposed an auxiliary self-distillation network for object segmentation, which features a multi-level pyramid representation branch to facilitate pixel-wise supervision.

The aforementioned methods do not address the issue of different granularities of features in hierarchical features, and overly tight constraints may lead to biased optimization directions. However, our approach combines a loosely constraint distribution vector-based method and tight constraints pixel-wise supervised method, allowing the model to be optimized with less bias during the self-knowledge distillation process, achieving superior segmentation performance.

\begin{figure*}[h]%
\centering
\includegraphics[width=1\textwidth]{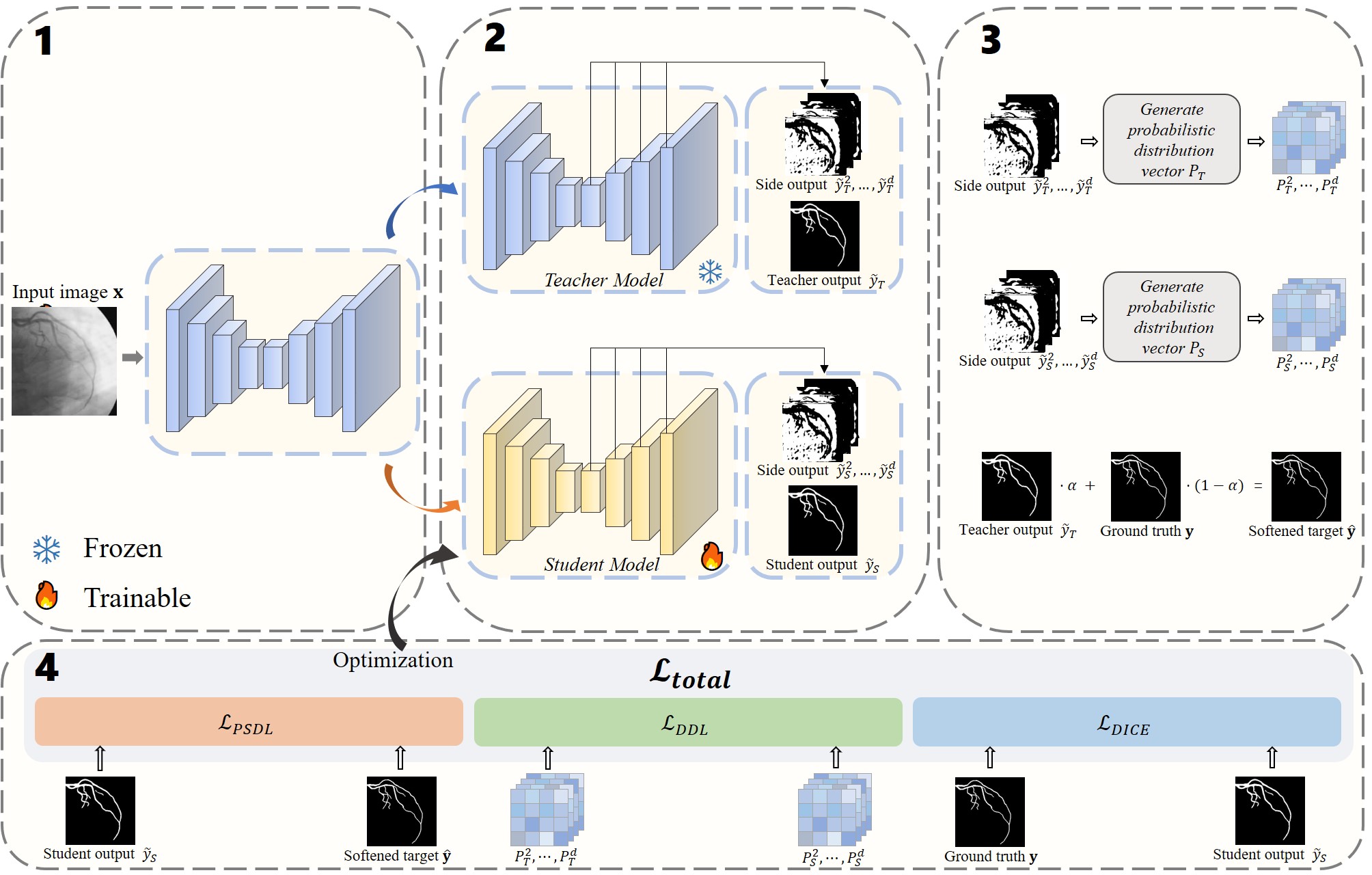}
\caption{The structure of Deep Self-knowledge Distillation.}\label{fig1}
\end{figure*}

\section{Methodology}\label{sec3}
\subsection{Preliminary}
In the coronary artery segmentation task, the input image, denoted as $\mathbf{x} \in \mathbb{R}^{H \times W}$, comprises pixels $x_{ij}$ where each pixel value $x_{ij} \in \mathbb{R}$ and indices $i$ and $j$ range from 1 to $H$ and $W$, respectively. The corresponding ground truth $\mathbf{y} \in \mathbb{R}^{H \times W}$ consists of binary pixel values $y_{ij} \in \{0,1\}$. The objective of training the segmentation network $N$, is to generate predictions $\hat{\mathbf{x}} = N(\mathbf{x})\approx \mathbf{y}$ effectively, where $\hat{\mathbf{x}}$ denotes the model's predicted output.

In segmentation networks employing an encoder-decoder architecture (e.g., U-Net \citep{ronneberger2015u}), the decoder comprises a series of up-sampling and convolutional layers. Consequently, each layer in the network hosts feature maps of different dimensions, represented as $X_{\mathrm{De}}^{k} \in \mathbb{R}^{\frac{H}{2^{k-1}} \times \frac{W}{2^{k-1}}}$, where $k \in \{1, \dots, d\}$, and $d$ denotes the depth of the network. Notably, $X_{\mathrm{De}}^{1} \in \mathbb{R}^{H \times W}$ represents the feature map at the shallowest layer of the network, which is parallel to the input image.

\subsection{Deep Self-knowledge Distillation}
Inspired by self-knowledge distillation \citep{kim2020self} and deep supervision \citep{zhou2018unet++}, we propose to hierarchically supervise feature maps at varying depths, while concurrently incorporate both loosely and tight constraints to facilitate self-knowledge distillation. The structure of our proposed Deep Self-knowledge Distillation is illustrated in Figure~\ref{fig1}.

\subsubsection{Deep Distribution Loss}
\begin{figure*}[h]%
\centering
\includegraphics[width=1\textwidth]{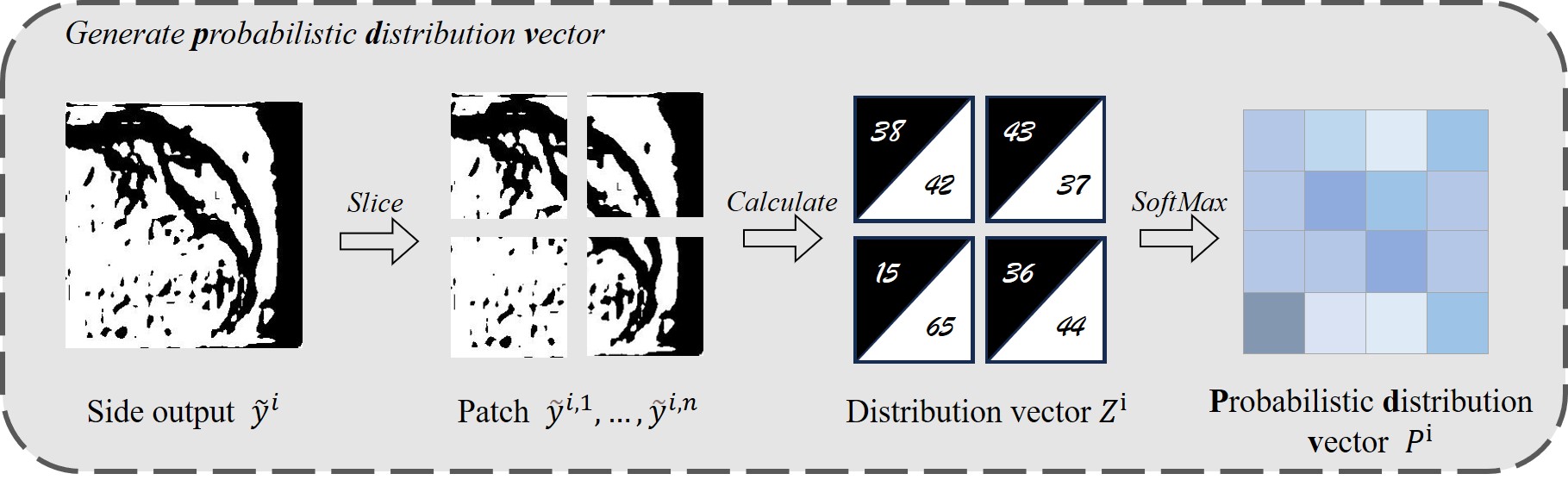}
\caption{The workflow of processing the side outputs to probabilistic distribution vectors.}\label{fig2}
\end{figure*}

Motivated by deep supervision as applied in UNet++ \citep{zhou2018unet++} and UNet 3+ \citep{huang2020unet}, we propose Deep Distribution Supervision. At each stage of the decoder, we obtain their side outputs, and supervise the differences in distribution between the side outputs of the student model and the teacher model via KL divergence. We anticipate that this hierarchical supervision through distillation will enable the student model to learn the representations of different granularity at different stages from the teacher model. The reason we opt for coarse-grained supervision using distribution vectors, rather than fine-grained pixel-wise supervision for side outputs, is that side feature maps at different layers learn features of varying granularity. Applying tight constraints directly to feature maps of varying granularity can easily introduce excessive inductive biases, leading to poor segmentation outcomes. Therefore, employing distribution vectors allows for model optimization in a direction with more relaxed constraints. Additionally, this method of supervising distributional differences, as opposed to supervising at a full-pixel level, may enhance generalizability and make the model more robustly optimized. The workflow of processing the side outputs to probabilistic distribution vectors is shown in Figure~\ref{fig2}. 

Given a decoder, we have side feature maps $X_{\mathrm{De}}^{i}$ from different layers, where the depth of the current layer $i\in\{1,\ldots,d\}$ and $d$ denotes the total depth of the model. These side feature maps are first fed into a $3 \times 3$ convolutional layer for dimension reduction, and then restored to the original input size via bilinear up-sampling $\mathcal{B}(X_{\mathrm{De}}^{k}; k-1):\mathbb{R}^{a\times b}\rightarrow \mathbb{R}^{a\cdot2^{k-1}\times b\cdot2^{k-1}}$, where $k$ represents the depth of the current layer. Since the side feature map parallel to the input image is the shallowest layer, it is denoted as $X_{\mathrm{De}}^{1}$. Finally, through the Sigmoid function, we can obtain the side outputs $\dot{X}_{\mathrm{De}}^{i}$ that match the shape of the ground truth at different stages of the decoder. The side outputs $\dot{X}_{\mathrm{De}}^{i}$ is defined as:
\begin{equation}
    \dot{X}_{\mathrm{De}}^{i}=\sigma(\mathcal{B}(f(X_{\mathrm{De}}^{i}); i-1))
\end{equation}
where the $\sigma(\cdot)$ denotes the Sigmoid activation function, $f(\cdot)$ denotes the $3 \times 3$ transition convolution for dimension reduction.

We now denote $\tilde{y}^{i}=\dot{X}_{\mathrm{De}}^{i}$ to represent the output of layer $i$, where $\tilde{y}^{i}$ takes values in 0 and 1 and $\tilde{y}^{i} \in \mathbb{R}^{H \times W}$. Subsequently, we can segment the output into individual patches and count the number of 0 and 1 in each patch, thus representing the distribution vector $\tilde{Z}^{i}$ for different layers. The distribution vector $\tilde{Z}^{i} \in \mathbb{R}^{n \times 2}$ is defined as:
\begin{equation}
    \tilde{Z}^{i}[m]=[\prod_{j=1}^{s}\prod_{k=1}^{s}\mathds{1}(\tilde{y}^{i,m}[j][k]=1),\prod_{j=1}^{s}\prod_{k=1}^{s}\mathds{1}(\tilde{y}^{i,m}[j][k]=0)]
\end{equation}
where $s$ denotes the patch size of the side output, $\tilde{Z}^{i}[m]$ denotes the $m^{th}$ row of the distribution vector $\tilde{Z}^{i}$ where $m \in \{1,\ldots, n\}$. The number of patches $n$ is equal to $H\times W/s^{2}$. $\tilde{y}^{i,m}$ denotes the $m^{th}$ patch of the side output $\tilde{y}^{i}$. $\mathds{1}(\cdot)$ is the indicator function.

By obtaining the distribution vector $\tilde{Z}^{i}$, we can derive a probabilistic distribution vector $P^{i}=[p_{1,0}^{i},p_{1,1}^{i},\ldots,p_{n,0}^{i},p_{n,1}^{i}]$ by softmax function. Meanwhile, we utilize the temperature scaling to soften these probabilities for better distillation \citep{hinton2015distilling}. We use the probabilistic distribution vector $P^{i}$ to represent the distribution of foreground and background in each side output $\tilde{y}^{i}$. The element $p_{j,k}^{i}$ in the probabilistic distribution vector $P^{i}$ is defined as:
\begin{equation}
    p_{j,k}^{i}=\frac{\exp(\tilde{Z}^{i}[j][k]/\tau)}{\sum_{u=1}^{n}\sum_{v=0}^{1}\exp(\tilde{Z}^{i}[u][v]/\tau)}
\end{equation}
where $\tau$ is the temperature scaling factor to smooth the probability distribution for better distillation. We obtain the probabilistic distribution vector $P^{i}_{S}$ from the student model as well as $P^{i}_{T}$ from the teacher model. Aiming to guide the student model in learning hierarchical representations from the decoder via the teacher model, we can use the KL divergence to optimize the student model, given by:
\begin{equation}
    \mathcal{L}_{DDL}=\sum_{i=1}^{d}KL(P_{S}^{i}||P_{T}^{i})=\sum_{i=1}^{d}(\sum_{j} P_{S}^{i}[j]\cdot log\frac{P_{S}^{i}[j]}{P_{T}^{i}[j]})
\end{equation}

\subsubsection{Pixel-wise Self-knowledge Distillation Loss}
Self-knowledge distillation enhances network performance by distilling its own knowledge, guiding the student model to learn representations from a superior performing teacher model. In other words, the teacher model is a past iteration of the student model, evolving throughout the training process, thus providing more precise supervision for guiding the student model. Previous work typically applied self-knowledge distillation to classification tasks. In this paper, we introduce a Pixel-wise Self-knowledge Distillation Loss for segmentation tasks, which supervises the student model's output by forming a smooth label through a linear combination of the teacher model's output and the ground truth. This approach not only improves the generalization ability of the network but also enhances the model's segmentation performance.

In the student model, we can obtain the output $\tilde{y}_{S}$ from the shallowest layer as the final prediction output, and similarly, we can obtain the output $\tilde{y}_{T}$ from the teacher model. For example, when the epoch is $t$, the current model is the student model. We use the model of epoch $t-1$ as the teacher model. By linearly combining the teacher model’s output $\tilde{y}_{T}$ with the ground truth $\mathbf{y}$, we can obtain a softened label $\hat{\mathbf{y}}$:
\begin{equation}
    \hat{\mathbf{y}}=\alpha \cdot {\tilde{y}}_{T}+(1-\alpha) \cdot \mathbf{y}
\end{equation}
where $\alpha$ is a hyper-parameter to adjust the proportion of the linear combination. The softened label integrates the knowledge from the teacher model and the hard labels of the ground truth, enhancing the generalization ability of the student model. Considering that in segmentation tasks we aim to achieve pixel-level classification, the supervision of the student model's output also needs to be pixel-wise. We aim for the predictions of the student model for each pixel to be as close as possible to the softened label $\hat{\mathbf{y}}$. Therefore, we use cross entropy loss to optimize the model. Our Pixel-wise Self-knowledge Distillation Loss is defined as follows:
\begin{equation}
    \mathcal{L}_{PSDL}=CE(\tilde{y}_{S}, \hat{\mathbf{y}})
\end{equation}

Where the $CE(\cdot)$ is the cross entropy loss.In the Self-knowledge Distillation process, it is imperative to take into account the evolving reliability of the teacher model, especially during the initial training phase when the model's understanding of the data is typically limited. To address this, we strategically adjust the value of the hyper-parameter $\alpha$ over time. Similar to learning rate scheduling techniques such as step-wise increases, exponential growth, or linear progression, we opt for a linear increase approach to minimize the complexity of hyper-parameter tuning. Consequently, the hyper-parameter $\alpha$ at the $t^{th}$ epoch is determined by the following formula:
\begin{equation}
    \alpha_t=\alpha_T\times\frac{t}{T}
\end{equation}
where $T$ denotes the number of training epochs, the $\alpha_T$ denotes the hyper-parameter $\alpha$ in the final epoch. 

\subsection{Full Optimization Objective}
In addition to the aforementioned Deep Distribution Loss and Pixel-wise Self-knowledge Distillation Loss, we also introduce dice loss which is widely used in segmentation tasks to balance the positive and negative samples in the image. Dice loss directly supervises the output of the student model against the ground truth, and it is defined as follows:
\begin{equation}
    \mathcal{L}_{DICE}=1-\frac{2\cdot \sum_{i=1}^{H}\sum_{j=1}^{W}\tilde{y}_{ij} \cdot y_{ij}}{\sum_{i=1}^{H}\sum_{j=1}^{W}\tilde{y}_{ij}+\sum_{i=1}^{H}\sum_{j=1}^{W}{y}_{ij}}
\end{equation}
where $\tilde{y}$ is the output of the student model. To summarize, our full optimization objective is defined by:
\begin{equation}
    \mathcal{L}_{total}=\mathcal{L}_{DDL}+\mathcal{L}_{PSDL}+\mathcal{L}_{DICE}
\end{equation}

The entire process of Deep Self-knowledge Distillation is presented in Algorithm~\ref{algo1}. 

\begin{algorithm}
    \caption{Deep Self-knowledge Distillation}\label{algo1}
    \KwIn{the input image $\mathbf{x}$, the ground truth $\mathbf{y}$, the original segmentation model $\mathbf{M}$ and the number of epoch $T$}
    \KwOut{the final version of the model $\mathbf{M}_{final}$}

    set $\mathcal{L}_{total} = \Phi, \mathbf{M}_{Teacher}=\Phi,\mathbf{M}_{Student}=\mathbf{M}$\;
    
    \For{$t = 1; t \le T$}
    {
        set $\mathbf{M}_{Student}=\mathbf{M}_{t}$
        
        compute $\tilde{y}_{S}=\mathbf{M}_{Student}(\mathbf{x})$\;

        compute $\mathcal{L}_{DICE}$   
        
        \eIf{$t=1$}
        {
            $\mathcal{L}_{total}=\mathcal{L}_{DICE}$
            
        }
        {
            compute $\tilde{y}_{T}=\mathbf{M}_{Teacher}(\mathbf{x})$ \;

            obtain probabilistic distribution vector $P_{S}$,  $P_{T}$ via $\tilde{y}_{S}$, $\tilde{y}_{T}$
        
            compute $\mathcal{L}_{DDL}$ and $\mathcal{L}_{PSDL}$

            $\mathcal{L}_{total}=\mathcal{L}_{DDL}+\mathcal{L}_{PSDL}+\mathcal{L}_{DICE}$\;
        }

        update $\mathbf{M}_{Teacher}=\mathbf{M}_{t}$ 

        optimize the student model $\mathbf{M}_{Student}$ with the total loss $\mathcal{L}_{total}$
    }
    \Return $\mathbf{M}_{final}$\;
\end{algorithm}

\section{Experiment}\label{sec4}

\subsection{Experimental setups}

\subsubsection{Datasets}
In this study, we employ the XCAD \citep{ma2021self} and DCA1 \citep{cervantes2019automatic} datasets to validate our approach. The X-ray angiography coronary artery disease (XCAD) dataset is obtained during stent implantation process using a General Electric Innova IGS 520 system. It consists of 1,621 coronary angiography images, each with a background mask frame annotated by experienced radiologists, and is of a resolution of $512 \times 512$. The DCA1 dataset comprises 134 grayscale images in portable gray map (PGM) format with a resolution of $300 \times 300$. DCA1 dataset has been annotated by expert cardiologists from the Cardiology Department of the Mexican Social Security Institute, UMAE T1-León. For both datasets, we have partitioned the data into training, validation, and testing sets following a 7:1:2 ratio, respectively.

Our proposed method is implemented using the PyTorch 1.7.0 framework and trained on an NVIDIA Tesla V100 SXM2 GPU with 32 GB of memory. The AdamW optimizer with a weight decay of 1e-5 is employed for the training of the segmentation networks. Additionally, we set the number of epochs to 100, the batch size to 4 with an initial learning rate of 0.001, and employ a step learning rate adjustment strategy that reduces the learning rate to 30\% of its initial value every 10 epochs.

We employ the Dice Similarity Coefficient (DSC) \citep{cardenas2018deep}, Accuracy (ACC), Sensitivity (SEN), and Intersection Over Union (IOU) as evaluation metrics to validate our model. Their definitions are as follows:
\begin{equation}
DSC=\dfrac{2|A\cap B|}{|A|+|B|}=\dfrac{2TP}{2TP+FP+FN}
\end{equation}\label{eq12}
\begin{equation}
ACC=\frac{TP+TN}{TP+TN+FP+FN}
\end{equation}\label{eq13}
\begin{equation}
SEN=\frac{TP}{TP+FN}
\end{equation}\label{eq14}
\begin{equation}
IOU=\frac{|A\cap B|}{|A\cup B|}=\frac{TP}{TP+FP+FN}
\end{equation}\label{eq15}
where TP, TN, FP, and FN denote the number of true positives, true negatives, false positives, and false negatives. A denotes the prediction and B denotes the ground truth.

\subsection{Comparison study}
To demonstrate the effectiveness of our proposed method, comparative experiments are conducted on several mainstream medical image segmentation models, including U-Net \citep{ronneberger2015u}, Attention U-Net \citep{oktay2018attention}, ResUnet \citep{zhang2018road}, U-Net++ \citep{zhou2018unet++}, HRNet \citep{sun2019deep}, Swin-unet \citep{cao2022swin}, DconnNet \citep{yang2023directional}, AtTransUNet \citep{li2023attransunet} and CT-Net \citep{zhang2024ct}. Additionally, U-Net3+ \citep{huang2020unet} is used as the baseline to evaluate the effectiveness of various components within the deep self-knowledge distillation framework. The experimental results are presented in Table~\ref{tab1}.

\begin{table*}[ht]
\caption{Comparative experimental results with other models and ablation study results on Deep Self-Knowledge Distillation Loss components, the best experimental results are shown in bold.}
\vspace{15pt}
\renewcommand\arraystretch{1.0}
\centering
\begin{tabular}{ccccc|cccc}
\toprule
\multirow{2}{*}{Method} & \multicolumn{4}{c}{XCAD}      & \multicolumn{4}{c}{DCA1}       \\
\cmidrule(lr){2-5} \cmidrule(lr){6-9}

                        & DSC (\%)   & ACC (\%)  & SEN (\%)  & IOU (\%)  & DSC (\%)  & ACC (\%)  & SEN (\%)  & IOU (\%)  \\
                        \midrule
U-Net                   & 77.53 & 97.43 & 75.75 & 64.32 & 76.08 & 97.12 & 75.28 & 62.83 \\
Attention U-Net         & 78.29 & 97.52 & 78.19 & 65.17 & 78.55 & 97.36 & 80.06 & 64.93 \\
ResUnet                 & 75.46 & 97.09 & 76.40 & 60.82 & 76.38 & 97.15 & 76.08 & 63.01 \\
U-Net++                 & 79.06 & 97.59 & 77.28 & 65.59 & 76.92 & 97.08 & 77.59 & 62.98 \\
HRNet                   & 77.32 & 97.32 & 76.92 & 64.89 & 77.81 & 97.23 & 79.29 & 64.03 \\
Swin-unet               & 74.18 & 96.92 & 76.19 & 58.99 & 76.21 & 96.73 & 76.29 & 62.34 \\
DconnNet                & 79.15 & 97.55 & 80.27 & 65.77 & 78.84 & 97.39 & \textbf{83.24} & 65.29 \\
AtTransUNet             & 78.92 & 97.41 & 79.11 & 65.34 & 77.67 & 97.24 & 80.34 & 64.89 \\
CT-Net                  & 78.71 & 97.49 & 78.93 & 64.97 & 76.91 & 97.08 & 80.67 & 65.09\\ \midrule
U-Net3+                 & 78.83 & 97.52 & 78.01 & 65.36 & 77.76 & 97.20 & 79.82 & 63.93 \\
U-Net3+ w/ $\mathcal{L}_{DDL}$  & 79.56 & 97.61 & 79.06 & 65.61 & 79.15 & 97.43 & 81.14 & 64.65 \\
U-Net3+ w/ $\mathcal{L}_{PSDL}$  & 80.02 & 97.65 & 79.78 & 65.85& 79.81 & 97.52 & 81.27 & 65.04 \\
U-Net3+ w/ $\mathcal{L}_{total}$  & \textbf{80.88} & \textbf{97.72} & \textbf{81.02} & \textbf{66.03} & \textbf{81.06} & \textbf{97.85} & 81.36 & \textbf{66.95} \\
\bottomrule
\end{tabular}
\label{tab1}
\end{table*}

From Table~\ref{tab1}, it is evident that our method achieved a certain degree of improvement on the XCAD dataset for the evaluation metrics of Dice Similarity Coefficient (DSC), Accuracy (ACC), Sensitivity (SEN), and Intersection Over Union (IOU). Specifically, our method's DSC improved by 2.19\% relative to the second-best comparative model and by 2.60\% relative to the baseline U-Net3+. Overall, our model demonstrated performance metrics of DSC, ACC, SEN, and IOU as 80.88\%, 97.72\%, 81.02\%, and 66.03\%, respectively, on the XCAD dataset. On the DCA1 dataset, our model also achieved desirable results in comparison. The DSC of our method improved by 2.82\% relative to the second-best comparative model and by 4.24\% relative to the baseline U-Net3+. Our model's performance on the DCA1 dataset was characterized by DSC, ACC, SEN, and IOU of 81.06\%, 97.85\%, 81.36\%, and 66.95\%, respectively.

\begin{figure*}[htbp]%
\centering
\includegraphics[width=1\textwidth]{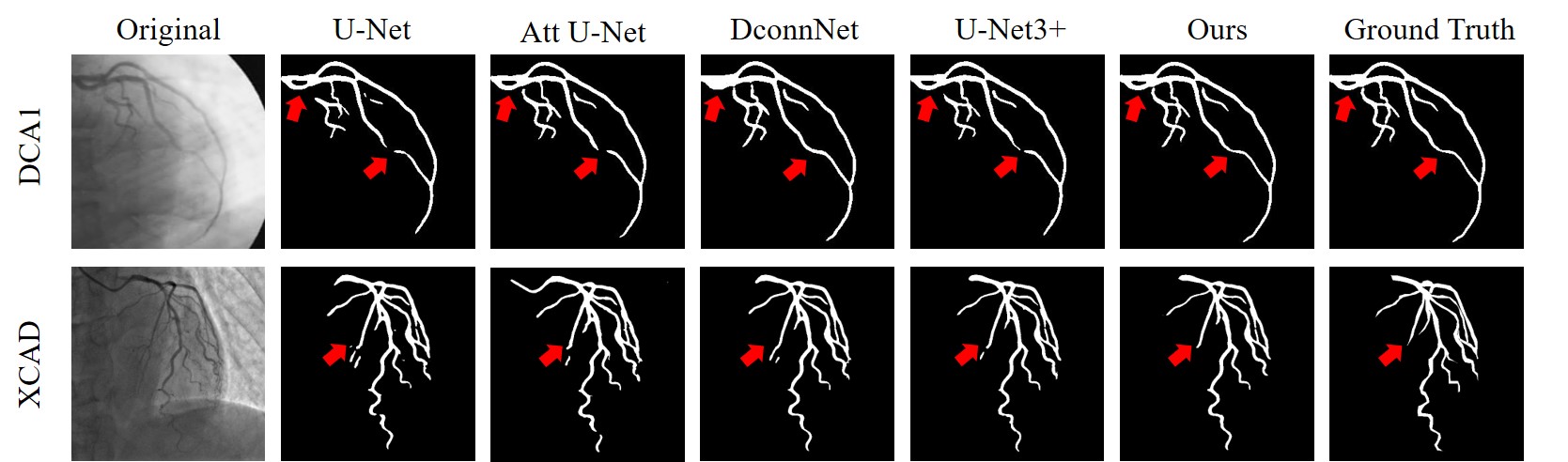}
\caption{Quantitative analysis of our proposed method compared to other comparative models.}\label{fig5}
\end{figure*}

Figure~\ref{fig5} illustrates the quantitative analysis of our proposed method compared to other comparative models. From the visual results, it can be observed that models such as U-Net and Attention U-Net (Att U-Net) exhibit poorer performance in evaluation metrics, which also translates to inadequate segmentation accuracy for finer vessels and insufficient vascular connectivity in the images. DconnNet, which focuses on improving segmentation connectivity, does enhance the segmentation accuracy for finer vessels; however, its pursuit of connectivity may lead to false positive segmentations. Our proposed method is closest to the ground truth in comparison, demonstrating the effectiveness and robustness of our approach.

\subsection{Ablation study}
In this section, we conduct an ablation study on the components of the Deep Self-Knowledge Distillation Loss, which includes the Deep Distribution Loss $\mathcal{L}_{DDL}$ and the Pixel-wise Self-Knowledge Distillation Loss $\mathcal{L}_{PSKL}$. We also examine the effects of the temperature scaling factor $\tau$ for smoothing the probability distributions, the number of patches $n$, and the hyper-parameter $\alpha$ that adjusts the linear combination ratio of the teacher model's output to the ground truth.

\subsubsection{Effects of different components}
From Table~\ref{tab1}, it is evident that the incorporation of both the Deep Distribution Loss and the Pixel-wise Self-Knowledge Distillation Loss has a positive impact on enhancing the baseline model's performance. The Deep Distribution Loss effectively transfers more domain-specific features from the teacher model to the student model by leveraging the knowledge contained in the side outputs. The Pixel-wise Self-Knowledge Distillation Loss contributes a degree of regularization by generating a softened target through the linear combination of the teacher model's output with the ground truth, thereby enhancing the model's robustness and generalization capabilities. Overall, the combination of Deep Distribution Loss and Pixel-wise Self-Knowledge Distillation Loss positively improves the final outcome of the model, enhancing the segmentation performance and increasing the credibility of the automated coronary artery segmentation model.

\subsubsection{Effects of temperature scaling factor $\tau$}
The ablation study results for the temperature scaling factor $\tau$ in the DCA1 dataset are presented in Table~\ref{tab2}. The temperature scaling factor $\tau$ is utilized to smooth the probability distribution output by the softmax function. A larger $\tau$ leads to a smoother output probability distribution, while a smaller 
$\tau$ results in a more concentrated distribution. In our experiments, we found that the optimal performance of the model occurred when $\tau=3$. Consequently, for the experiments mentioned in Table~\ref{tab1}, we set the temperature scaling factor $\tau$ to 3.

\begin{table}[!htbp]
\renewcommand\arraystretch{1.0}
\caption{Ablation study results on temperature scaling factor $\tau$, the best results are bond.}
\vspace{15pt}
\centering
\begin{tabular}{ccccc}
\toprule
\multirow{1}*{$\tau$}  & \multicolumn{1}{c}{DSC (\%)}  & \multicolumn{1}{c}{ACC (\%)} & \multicolumn{1}{c}{SEN (\%)}  & \multicolumn{1}{c}{IOU (\%)} \\
\midrule
0.5  & 78.96 & 96.93 & 79.26 & 65.26 \\
1    & 79.15 & 97.34 & 79.47 & 65.81 \\
2    & 80.53 & 97.56 & 80.86 & 66.05 \\
3    &\textbf{81.06} & \textbf{97.85} & \textbf{81.36} & \textbf{66.95} \\
4    & 79.38 & 97.65 & 79.34 & 66.23 \\
5    & 79.07 & 97.40 & 79.10 & 65.84 \\
\bottomrule
\end{tabular}
\label{tab2}
\end{table}

\subsubsection{Effects of patch numbers $n$}
The ablation study results for the patch numbers $n$ in the DCA1 dataset are shown in Table~\ref{tab3}. Patch numbers $n$ determine the granularity of the probabilistic distribution vector. A larger $n$ results in a finer granularity of the probabilistic distribution vector, imposing a tighter constraint on the distillation process. Our experiments revealed that the optimal performance was achieved when $n=4\times4=16$. Therefore, for the experiments referenced in Table~\ref{tab1}, we set the patch numbers $n$ to 16.

\begin{table}[htbp]
\renewcommand\arraystretch{1.0}
\caption{Ablation study results on patch numbers $n$, the best results are bond.}
\vspace{15pt}
\centering
\begin{tabular}{ccccc}
\toprule
\multirow{1}*{$n$}  & \multicolumn{1}{c}{DSC (\%)}  & \multicolumn{1}{c}{ACC (\%)} & \multicolumn{1}{c}{SEN (\%)}  & \multicolumn{1}{c}{IOU (\%)} \\
\midrule
1  & 77.96 & 96.73 & 77.98 & 64.51 \\
4  & 79.65 & 97.44 & 80.18 & 66.45 \\
16  &\textbf{81.06} & \textbf{97.85} & \textbf{81.36} & \textbf{66.95} \\
64   & 79.89 & 97.47 & 80.46 & 66.01 \\
\bottomrule
\end{tabular}
\label{tab3}
\end{table}

\subsubsection{Effects of soften adjustment parameter $\alpha$}
The ablation study results for the soften adjustment parameter $\alpha$ in the DCA1 dataset are depicted in Table~\ref{tab4}. The soften adjustment parameter $\alpha$ adjusts the ratio of the teacher model's output to the ground truth in the softened target. A larger $\alpha$ indicates that the proportion of the teacher model's output in the softened target increases as the model trains. We found that the optimal performance was attained when $\alpha=0.5$. An excessively large $\alpha$ can introduce too much bias into the pixel-level distillation loss during the later stages of model training due to the high proportion of the teacher model's output, leading to suboptimal segmentation results. Conversely, a too small 
$\alpha$ implies that the softened target does not provide an adequate amount of regularization, resulting in insufficient generalization capability and poor model performance. Thus, for the experiments detailed in Table~\ref{tab1}, we set the soften adjustment parameter $\alpha$ to 0.5.

\begin{table}[htbp]
\renewcommand\arraystretch{1.0}
\caption{Ablation study results on soften adjustment parameter $\alpha$, the best results are bond.}
\vspace{15pt}
\centering
\begin{tabular}{ccccc}
\toprule
\multirow{1}*{$\alpha$}  & \multicolumn{1}{c}{DSC (\%)}  & \multicolumn{1}{c}{ACC (\%)} & \multicolumn{1}{c}{SEN (\%)}  & \multicolumn{1}{c}{IOU (\%)} \\
\midrule
0.1  & 78.56 & 97.32 & 78.88 & 66.01 \\
0.3  & 79.84 & 97.37 & 79.17 & 66.25 \\
0.5  &\textbf{81.06} & \textbf{97.85} & 81.36 & \textbf{66.95} \\
0.7   & 80.85 & 97.81 & \textbf{81.43} & 66.87 \\
0.9   & 76.07 & 96.23 & 77.06 & 62.27 \\
\bottomrule
\end{tabular}
\label{tab4}
\end{table}

\section{Conclusion}\label{sec5}
In this paper, we introduce Deep Self-knowledge Distillation, a self-distillation method that employs hierarchical outputs for supervision. The deep self-knowledge distillation loss is composed of two parts: the Deep Distribution Loss and the Pixel-wise Self-knowledge Distillation Loss. The Deep Distribution Loss represents the similarity between the side outputs of the student and teacher models using probabilistic distribution vectors derived from the transformation of side feature maps in the encoder-decoder structured segmentation network, and it employs the Kullback-Leibler divergence to narrow the distance between these vectors. This loosely constrained approach effectively injects valid knowledge into the student model without introducing excessive bias. Moreover, it is at no computational cost during inference. The Pixel-wise Self-knowledge Distillation Loss linearly combines the teacher model's output with the ground truth to generate a softened target for pixel-wise supervision of the student model's output. This distillation process introduces regularization during the model's final output generation, enhancing the model's performance. Furthermore, we aim to explore additional methods in the future for extracting effective information from the teacher model to inject into the student model, thereby achieving superior segmentation performance.

\bibliography{ref}

\end{document}